\documentclass[12pt]{iopart}

\begin{document}

\title[]{A study of electronic structure of FeSe$_{1-x}$Te$_{x}$
chalcogenides by Fe and Se K-edge x-ray absorption near edge structure
measurements}

\author{B. Joseph$^{1}$, A. Iadecola$^{1}$, L. Simonelli$^{2}$, Y.
Mizuguchi$^{3,4}$, Y. Takano$^{3,4}$, T. Mizokawa$^{5,6}$, N. L.
Saini$^{1,6}$} 
\address{$^{1}$Dipartimento di Fisica, Universit\`{a}
di Roma ``La Sapienza", P. le Aldo Moro 2, 00185 Roma, Italy}
\address{$^{2}$European Synchrotron Radiation Facility, 6 RUE Jules
Horowitz BP 220 38043 Grenoble Cedex 9 France} 
\address{$^{3}$National Institute for Materials Science, 1-2-1 Sengen, 
Tsukuba 305-0047, Japan}
\address{$^{4}$JST-TRIP, 1-2-1 Sengen,Tsukuba 305-0047, Japan}
\address{$^{5}$Department of Physics, University of Tokyo, 5-1-5
Kashiwanoha, Kashiwa, Chiba 277-8561, Japan} 
\address{$^{6}$Department
of Complexity Science and Engineering, University of Tokyo, 5-1-5
Kashiwanoha,Kashiwa, Chiba 277-8561, Japan}

\begin{abstract}

Fe K-edge and Se K-edge x-ray absorption near edge structure (XANES)
measurements are used to study FeSe$_{1-x}$Te$_{x}$ electronic
structure of chalcogenides.  An intense Fe K-edge pre-edge peak due to
Fe 1s$\to$3d (and admixed Se/Te $p$ states) is observed, showing
substantial change with the Te substitution and X-ray polarization.
The main white line peak in the Se K-edge XANES due to Se 1s $\to$ 4p
transition appear similar to the one expected for Se$^{2-}$ systems
and changes with the Te substitution.  Polarization dependence reveals
that unoccupied Se orbitals near the Fermi level have predominant
$p_{x,y}$ character.  The results provide key information on the
hybridization of Fe $3d$ and chalcogen $p$ states in the Fe-based
chalcogenide superconductors.

\end{abstract}


\maketitle

\section{Introduction}

Observation of superconductivity in the FeSe$_{1-x}$Te$_{x}$
chalcogenides with interplaying superconductivity and magnetism
\cite{HsuPNAS08,FangPRB08,YehEPL08,MedNatMat09,KhasaPRB09} is of high
interest due to apparent simplicity of these materials in comparison
to the Fe-based pnictides \cite{HosJPSJ09,RenAdMat09}.  Indeed, the
structure of FeSe$_{1-x}$Te$_{x}$ contains a simple stacking of edge
sharing Fe(Se,Te)$_{4}$ tetrahedra \cite{HsuPNAS08,FangPRB08,YehEPL08}
without any sandwiching spacer \cite{RicciSUST10}.  The
superconductivity is very sensitive to defects and disorder
\cite{McQPRBPRL09}, with the T$_{c}$ in the ternary
FeSe$_{1-x}$Te$_{x}$ system increasing up to a maximum of 14 K
\cite{FangPRB08,YehEPL08,HoriJPSC09}.  However, the missing spacer
layer in the FeSe$_{1-x}$Te$_{x}$ might be a cause of locally broken
symmetry in the ternary system
\cite{JosephPRB10,IadEPL10,LoucaPRB10,TegelSSC10} with the Se and Te
occupying distinct sites.  Although, the electronic states near the
Fermi level in the FeSe$_{1-x}$Te$_{x}$ chalcogenides are given by the
five Fe 3d-orbitals, it is also true that the fundamental transport
properties are very sensitive to the chalcogen (Se/Te) height from the
Fe-Fe sublattice \cite{MiyakeJPSJ10,MoonPRL10,SubediPRB09}, indicating
importance of interaction of Fe 3d states with the chalcogen orbitals.
Therefore, it is of key importance to study details on the electronic
structure of the Fe 3d and the interacting chalcogen orbitals.

X-ray absorption near-edge structure (XANES) spectroscopy is a site
specific probe of distribution of the valence electrons and local
chemistry, with the final states in the continuum being due to
multiple scattering resonances of the photoelectron in a finite
cluster \cite{Konings}.  Unlike photoemission experiments, there are
negligible surface effects (and multiplet effects), making it a very
useful finger print probe of unoccupied valence states and site
selective local chemistry.  The XANES spectroscopy has been already
used to study the electronic structure and the local geometry in the
Fe-based REFeAsO (RE=rare earth) oxypnictides
\cite{JosephCM,XuCMEPL,BondinoPRL,KrollPRB,Ignatov08,MustreJS09,ChangPRB09,ShenPRB}
and AFe$_{2}$As$_{2}$ (A=Ba,Sr,Ca) pnictides
\cite{ShenPRB,BondinoPRB}.  Indeed, important information on the
electronic correlations \cite{BondinoPRL,KrollPRB,ShenPRB} as well on
the local geometry \cite{JosephCM,XuCMEPL,Ignatov08} has been
obtained.  However, there is hardly any systematic XANES study on the
Fe-based chalcogenides.  Here, we have used XANES spectroscopy to
study electronic structure of unoccupied states in
FeSe$_{1-x}$Te$_{x}$ system.  A combined analysis of Fe and Se K-edge
XANES has permitted to uncover important features of the unoccupied
states near the Fermi level.  We find a substantial hybridization
between the Fe $3d$ and chalcogen $p$ states, getting redistributed
systematically with the Te substitution.  Furthermore, using polarized
XANES on single crystal sample, we have found that chalcogen $p_{xy}$
orbitals should be predominantly interacting with the Fe 3d orbitals
in the chalcogenides.  The results underline importance of the $p-d$
hybridization in the Fe-based chalcogenide superconductors.


\section{Experimental Details}

Unpolarized and polarized X-ray absorption measurements were made at
the beamline BM29 of the European synchrotron radiation facility
(ESRF), Grenoble, on the FeSe$_{1-x}$Te$_{x}$ samples characterized
for their superconducting and structural properties
\cite{MizJPSJ09,KidaJSPJ09}.  Unpolarized spectra were measured on
powder samples \cite{MizJPSJ09} of FeSe$_{1-x}$Te$_{x}$ (x=0.0, 0.5,
1.0) to explore the composition dependence, while single crystal
sample of FeSe$_{0.25}$Te$_{0.75}$ \cite{KidaJSPJ09} was selected to
study polarization dependence of the XANES features.  The synchrotron
radiation emitted by a bending magnet source at the 6 GeV ESRF storage
ring was monochromatized by a double crystal Si(311) monochromator and
sagittally focused on the samples, mounted in a continuous flow He
cryostat.  For the polarized measurements on the
FeSe$_{0.25}$Te$_{0.75}$ single crystal sample, normal incidence
geometry with linearly polarized light falling parallel to the
ab-plane and grazing incidence geometry with linearly polarized polarized 
light falling nearly perpendicular to the ab-plane were used.  The Se K-edge
absorption spectra were recorded by detecting the Se K$_{\alpha}$
fluorescence photons, while Fe K$_{\alpha}$ fluorescence photons were
collected over a large solid angle using multi-element Ge-detector for
measuring the Fe K-edge absorption.  For the unpolarized spectra on
the powder samples we used simultaneous detection of the fluorescence
signal and the transmission yield.  The sample temperature was
controlled and monitored within an accuracy of $\pm$1 K. As a routine
experimental approach, several absorption scans were collected to
ensure the reproducibility of the absorption spectra, in addition to
the high signal to noise ratio.  After subtracting linear pre-edge
background, the XANES spectra were normalized to the energy dependent
atomic absorption, estimated by a linear fit to the extended x-ray
absorption fine structure (EXAFS) region away from the absorption
edge.


\section{Results and Discussion}

\subsection{Fe K-edge XANES}

Figure 1 shows normalized Fe K-edge XANES spectra of
FeSe$_{1-x}$Te$_{x}$ (x=0.0, 0.5, 1.0) measured at T = 30 K. The
spectra are close to the one for a reference of Fe$^{2+}$ standard
\cite{Ignatov08,ChangPRB09}, consistent with the Fe$^{2+}$ state.  The
general features of the Fe K-edge are similar to those reported for
the Fe-based pnictides \cite{Ignatov08,MustreJS09,ChangPRB09}.  The
near edge features are marked with A, B, C, D and E. The K- edge
absorption process is mainly governed by the
\textit{1s$\to$$\epsilon$p} dipole transition and hence continuum
states with \textit{$\epsilon$p} symmetries (and admixed states) can
be reached in the final state.  In addition to the dipole transition,
a direct quadrupole transition in the unoccupied \textit{3d} states is
seen as pre-peak A, mixed with the dipole contribution (mixing with
the $4p$ states due to local distortions).  Therefore, apart from the
density of the unoccupied electronic states, a changing pre-edge
intensity can be an indicator of a changing local geometry or
distortion around the Fe atom.  Here, the pre-peak A ($\sim$7111 eV)
is due to the \textit{1s$\to$3d} quadrupole transition, with some
dipole contribution due to the admixed $p$ states.  The feature B
(shoulder structure of the main absorption jump at $\sim$7117 eV)
appears due to the \textit{1s$\to$4p} transition.  The peak like
structure C ($\sim$7120 eV) should be driven by the \textit{1s$\to$4p}
states admixed with the $d$ states of the chalcogen atoms.  A
significant change in the pre-peak A intensity can be seen as a
function of Te substitution with the one for the FeSe appearing more
intense (see e.g., the inset showing a zoom over the peak A).
Similarly, the feature C looks more intense for the FeSe sample due to
larger mixing of Fe $4p$/chalcogen $d$ states.  The features at higher
energies are mainly due to the photoelectron multiple scattering with
the nearest neighbours.

\begin{figure}
\input{epsf}
\epsfxsize 10.5 cm
\centerline{\epsfbox{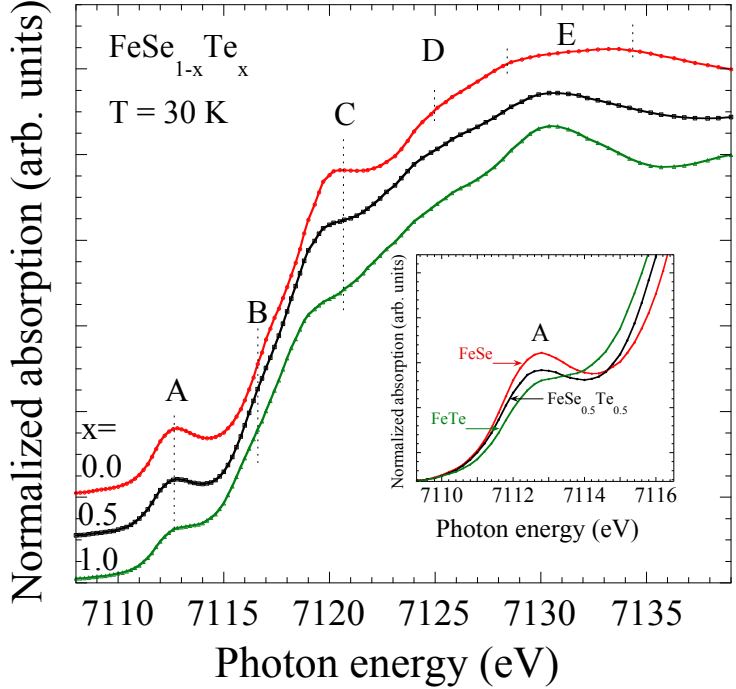}}
\caption{\label{fig:epsart}
Fe K-edge XANES spectra of FeSe$_{1-x}$Te$_{x}$ (x=0, 0.5, 1) measured
at 30 K. The inset shows a zoom over the near edge features A.}
\end{figure}

For a further clarification on the Fe K-edge spectral features, we
have studied polarization dependence of the XANES measured on a single
crystal of representative FeSe$_{0.25}$Te$_{0.75}$ sample.  Fig.  2.
shows the Fe K-edge XANES spectra with varying polarization.  A
significant polarization dependence of Fe K-edge features can be seen.
In particular, the pre-peak shows a significant increase from parallel
to the almost perpendicular polarization (E$\parallel$75 degree).
This indicates increased density of unoccupied Fe 3d states admixed
with the p states originated from the chalcogen atoms.  The
polarization dependence appears similar to the one found for the
oxypnictides \cite{ChangPRB09}.  The results also appear consistent
with the local-density approximation (LDA) calculations for these
materials \cite{MiyakeJPSJ10}.  Similarly, the peak C gets more
intense for the perpendicular geometry, mainly due to higher density
of states for Fe $4p$/chalcogen $d$ hybrid bands along the c-axis.  On
the other hand, the peak B appears hardly affected with the
polarization.  The polarization dependence can be understood also in
terms of different local geometry of the system in the the two
directions.  Indeed, with E$\parallel$ab plane, the Fe-Fe planar
orbitals are available for the transition while with E$\parallel$c,
the mixing of the chalcogen $p$ (peak A) and chalcogen $d$ (peak C) is
expected to be more prominent.

\begin{figure}
\input{epsf}
\epsfxsize 10.5cm
\centerline{\epsfbox{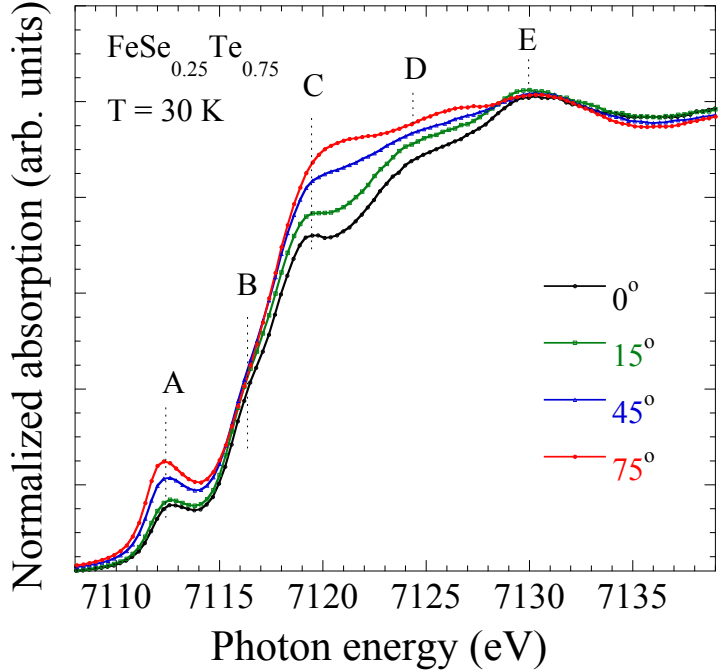}}
\caption{\label{fig:epsart}Polarized Fe K-edge XANES
spectra of FeSe$_{0.25}$Te$_{0.75}$ measured at 30 K. The
0$^\circ$ correspond to the E$\parallel$ab while the
75$^\circ$ represent almost E$\parallel$c case.  Strong
polarization effect is evident on the pre-peak A and the peak C.}
\end{figure}

Coming back to the substitution effect (Fig.  1), a clear energy shift
can be seen for the features B and the peak C. Indeed, the peak C
is shifted by almost 0.8 eV lower energy for the FeTe with respect to
the FeSe.  Since this peak is derived by the Fe-chalcogen orbitals
mixing, any change in the Fe-chalcogen bond length is expected to
influence the peak position, related by the
$\Delta$E$\propto$1/d$^{2}$ relation for a XANES resonance
\cite{BianPRB85}.  The Fe-chalcogen bondlength in the FeTe is
$\approx$ 2.6 $\AA$, larger than the Fe-chalcogen bondlength in the
FeSe ($\approx$ 2.4 $\AA$).  A gradual and substantial decrease of the
pre-peak (Fig.  1) derived by \textit{1s$\to$3d} quadrupole transition
and a dipole transition due to admixed chalcogen p states is
consistent with the longer Fe-Te distance.  On the other hand, an
apparent broadening and the shift of the peak C for the
FeSe$_{0.5}$Te$_{0.5}$ should be due to the local inhomogeneity of the
ternary system, charcaterized by coexisting Fe-chalcogen bondlengths
\cite{JosephPRB10,IadEPL10,LoucaPRB10,TegelSSC10}.  Similarly, the
peak B is shifted towards lower energy for the FeTe with respect to
the FeSe, merely due to the fact that the Fe-Fe bondlength for the
FeSe ($\approx$ 2.66 $\AA$) is lower than the one for the FeTe
($\approx$ 2.69 $\AA$).  Following the above arguments we can state
that the higher intensity of the pre-peak should be due to higher
mixing of the chalcogen p orbitals with the Fe $3d$ states consistent
with the shorter Fe-chalcogen bondlength, and hence the dipole
contribution appears to be changing, with higher number of available
unoccupied states for the transition from the Fe $1s$ states.  It
should be mentioned that a sophisticated theoretical model is required
for a quantitative estimation of the dipole and quadrupole
contribution in the present system with Fe in a tetrahedral geometry.


\subsection{Se K-edge XANES}

Figure 3 compares Se K-edge XANES spectra of FeSe and
FeSe$_{0.5}$Te$_{0.5}$ samples measured at T = 30 K. There are two
main features, (i) a sharp peak A ($\sim$12658 eV), which is mainly
due to a direct \textit{1s$\to$4p} dipole transition and, (ii) the
broad hump B (about 7 eV above the peak A), should be a multiple
scattering of the photoelectron with the near neighbours.  The spectra
are typical of Se$^{2-}$ systems \cite{SeK1,SeK2} with position of the
peak A in all the samples being consistent with earlier studies on
similar systems.  The peak A appears to have lower intensity for the
Te substituted samples, suggesting decreased number of unoccupied Se
$4p$ states near the Fermi level with Te substitution.  In addition,
the multiple scattering hump B shows some evident changes with the Te
substitution, mainly due to changing local geometry around the Se
atoms.  Indeed, this hump appears to be getting broader for the Te
substituted samples with an overall shift towards lower energy.  This
should be related to local inhomogeneity of the ternary system, seen
by EXAFS measurements \cite{JosephPRB10,IadEPL10}.

\begin{figure}
\input{epsf}
\epsfxsize 10.5cm
\centerline{\epsfbox{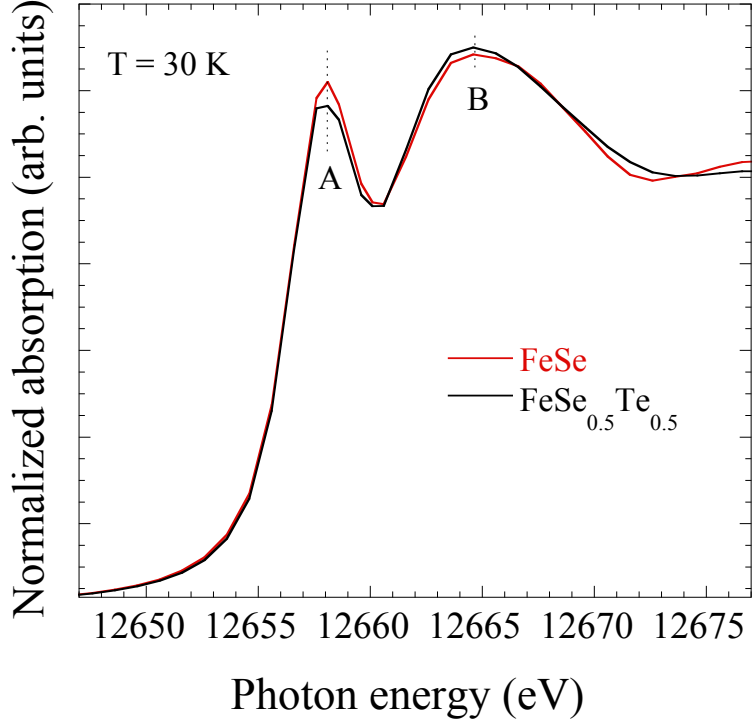}}
\caption{\label{fig:epsart} Se K-edge XANES spectra of
FeSe$_{1-x}$Te$_{x}$ (x=0, 0.5) measured at T = 30 K.}
\end{figure}

To have further details of the unoccupied Se electronic states, we
have measured the Se K-edge XANES in different polarizations on the
FeSe$_{0.25}$Te$_{0.75}$ single crystal sample.  Fig.  4 shows
normalized Se K-edge XANES measured with the polarization parallel and
nearly perpendicular to the ab-plane of the single crystal sample.
There is large polarization dependence, with the peak A due to the
\textit{1s$\to$4p} dipole transition appearing extremely damped in the
spectrum obtained using perpendicular polarization.  This merely
brings a conclusion that unoccupied Se $4p$ states near the Fermi
level in these systems are mainly derived by the p$_{x,y}$ symmetry.
Incidentally, the hump B also shows a strong polarization effect.
Indeed, the hump has very different spectral shape in the two
geometries, appearing with apparently two broad features in the
E$\parallel$c geometry unlike a single broad feature in the
E$\parallel$ab geometry.  The lower energy feature of the hump in the
E$\parallel$c geometry shows an overall shift with respect to the hump
in the E$\parallel$ab geometry.  Since the hump is due to multiple
scattering including Se-Fe, Se-Se and Se-Te shells, different spectral
shapes are likely to be due to different contributions of these
scattering paths in the two polarization geometries.  The fact that,
while Se-Fe paths are equally seen in the two geometries, the Se-Se/Te
paths are different with some of the scattering paths missing in the
E$\parallel$c geometry, and hence an apparent two peak spectral shape
in the E$\parallel$c geometry unlike a broader hump B for the
E$\parallel$ab polarization.  
However, shell by shell full multiple
scattering calculations need to be performed to study the details
of these local geometrical variations, which are beyond the
scope of the present paper’s focus on the electronic structure.
\begin{figure}
\input{epsf}
\epsfxsize 10.5cm
\centerline{\epsfbox{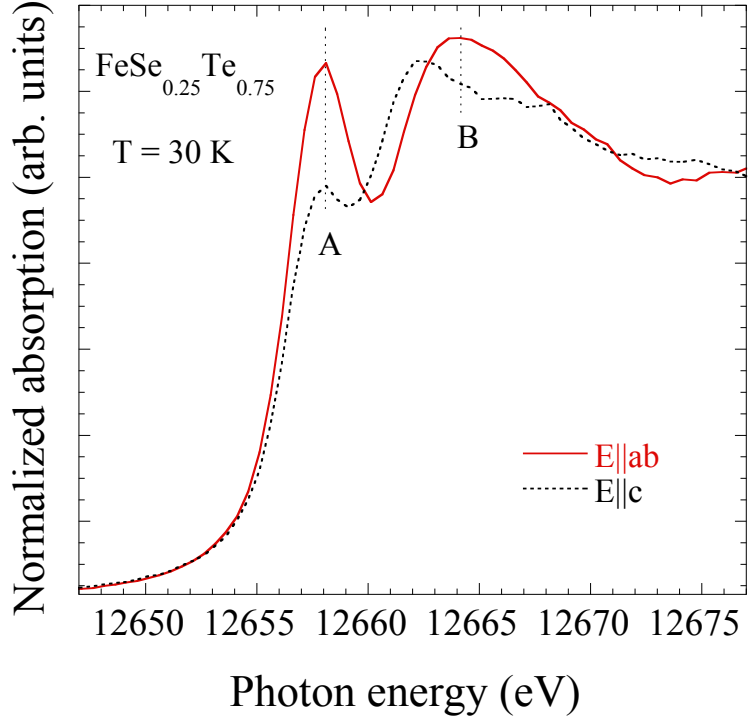}}
\caption{\label{fig:epsart}Polarized Se K-edge XANES spectra of
FeSe$_{0.25}$Te$_{0.75}$ crystal measured at T = 30 K with
polarization of the beam along the two high symmetry crystal axes.}
\end{figure}

\section{Summary}

In summary, we have studied electronic structure of
FeSe$_{1-x}$Te$_{x}$ chalcogenides by a combination of Fe and Se
K-edge X-ray absorption near edge structure spectroscopy.  From the Fe
K-edge data we find a gradual and substantial decrease of the pre-peak
derived by \textit{1s$\to$3d} quadrupole transition and a dipole
transition due to admixed chalcogen $p$ states.  The damping seems to
be due to lower mixing of chalcogen $p$ states in the Te containing
systems, consistent with the longer Fe-Te distance.  Again, the Se
K-edge XANES spectra reveal a damping of unoccupied $4p$ states,
consistent with the Fe K-edge XANES suggesting lower admixing of the
Fe $3d$ with the chalcogen $p$ states in the Te containing systems.
Furthermore, the polarized Se K-edge XANES reveals predominant
$p_{x,y}$ character of the chalcogen $p$ states that should be
involved in the admixing with the unoccupied Fe $3d$ states near the
Fermi level.  The results are consistent with strong chalcogen height
sensitivity of the fundamental electronic structure, that seems to be
related with redistribution of the admixed Fe 3$d$- chalcogen $p$
states.

\section*{Acknowledgments}

The authors thank the ESRF staff for the help and cooperation during
the experimental run.


\section*{References}
\begin{thebibliography}{0}

\bibitem{HsuPNAS08} Hsu, F.-C., Luo, J.-Y., Yeh, K.-W., Chen, T.-K., 
Huang, T.-W., Wu, P. M., Lee, Y.-C., Huang, Y.-L., Chu, Y.-Y., 
Yan, D.-C., Wu, M.-K., 
Proc. Nat. Acad. Sci. 105, 14262 (2008).\\

\bibitem{FangPRB08}
Fang M. H., Pham H. M., Qian B., Liu T. J., Vehstedt E. K., Liu Y., 
SpinuL., and Mao Z. Q., 
Phys. Rev. B 78, 224503 (2008). \\

\bibitem{YehEPL08} Yeh, K.-W., Huang, T.-W., Huang, Y.-l., 
Chen, T.-K., Hsu, F.-C., Wu, P. M., Lee, Y.-C., Chu, Y.-Y., 
Chen, C.-L., Luo, J.-Y., Yan, D.-C., Wu, M.-K., 
EPL 84, 37002 (2008).\\

\bibitem{MedNatMat09} Medvedev, S., McQueen, T. M., Troyan, I. A., 
Palasyuk, T., Eremets, M. I., Cava, R. J., Naghavi, S., 
Casper, F., Ksenofontov, V., Wortmann, G., Felser, C., 
Nat. Mat. 8, 630-633 (2009).\\

\bibitem{KhasaPRB09} Khasanov, R., Bendele, M., Amato, A., 
Babkevich, P., Boothroyd, A. T., Cervellino, A., Conder, K., 
Gvasaliya, S. N., Keller, H., Klauss, H. H., Luetkens, H., 
Pomjakushin, V., Pomjakushina, E., Roessli, B., 
Phys. Rev. B 80, 140511  (2009).\\

\bibitem{HosJPSJ09} Ishida K., Nakai Y., and Hosono H., 
J. Phys. Soc. Jpn. 78, 062001 (2009).\\

\bibitem{RenAdMat09} Ren Z.-A. and Zhao Z.-X., 
Advanced Materials 21, 4584 (2009), ISSN 1521-4095. \\

\bibitem{RicciSUST10}Ricci A., Joseph B., Poccia N., Xu W., Chen D.,
Chu W.S., Wu Z.Y., Marcelli A., Saini N.L. and Bianconi A., Supercond.
Sci.  Technol.  23, 052003 (2010).\\

\bibitem{McQPRBPRL09} McQueen, T. M., Williams, A. J.,
Stephens, P. W., Tao, J., Zhu, Y., Ksenofontov, V., Casper, F.,
Felser, C., Cava, R. J., Phys. Rev. Lett. 103, 057002
(2009);\\
McQueen T. M., Huang Q., Ksenofontov V., Felser C., Xu Q.,
Zandbergen H., Hor Y. S., Allred J., Williams A. J., Qu D., Checkelsky
J., Ong N. P., Cava R. J., Phys. Rev. B 79, 014522 (2009).\\

\bibitem{HoriJPSC09} Horigane, K., Hiraka, H., Ohoyama, K., 
J. Phys. Soc. Jpn. 78, 074718  (2009). \\

\bibitem{JosephPRB10}Joseph B., Iadecola A., Puri A., Simonelli L.,
Mizuguchi Y., Takano Y., and Saini N. L., Phys.  Rev.  B 82, 020502(R)
(2010).\\

\bibitem{IadEPL10}Iadecola A., Joseph B., Simonelli L., Mizuguchi Y.,
Takano Y. and Saini N. L., EPL, 90, 67008 (2010). \\

\bibitem{LoucaPRB10} Louca D., Horigane K., Llobet A., Arita R., Ji S.,
Katayama N., Konbu S., Nakamura K., Koo T.-Y., Tong P., and Yamada K.,
Phys.  Rev.  B 81, 134524 (2010).\\

\bibitem{TegelSSC10} Tegel M., Loehnert C., Johrendt D.,
Solid State Commun. 150, 383 (2010).\\

\bibitem{MiyakeJPSJ10} Miyake, T., Nakamura, K., Arita, R., Imada, M.,
J. Phys. Soc. Jpn. 79 044705 (2010).\\

\bibitem{MoonPRL10} Moon, C.-Y., Choi, H. J., Phys. Rev. Lett.
104, 057003 (2010).\\

\bibitem{SubediPRB09} Subedi, A., Zhang, L., Singh, D. J., Du, M. H., 
Phys. Rev. B 78, 134514 (2009).\\

\bibitem{Konings}Bianconi A., Dell'Ariccia M., Durham P. J. and 
Pendry J. B., Phys.  Rev.  B 26, 6502 (1982);\\ 
EXAFS and Near Edge Structure
edited by Bianconi A., Incoccia L., and Stipcich S. (Springer-Verlag,
Berlin, 1982);\\ 
Bianconi A. in X-ray Absorption: Principles,
Applications, Techniques of EXAFS, SEXAFS, XANES, edited by Prins R.
and Koningsberger D. C. (Wiley, New York, 1988).\\

\bibitem{JosephCM} Joseph B., Iadecola A., Fratini M., Bianconi A.,
Marcelli A. and Saini N. L., J. Phys.: Condens.  Matter 21 432201
(2009).\\

\bibitem{XuCMEPL} Xu W., Marcelli A., Joseph B., Iadecola A., Chu W.
S., Di Gioacchino D., Bianconi A., Wu Z. Y. and Saini N. L., J. Phys.:
Condens.  Matter 22 125701 (2010);\\ 
Xu W., Joseph B., Iadecola A.,
Marcelli A., Chu W. S., Di Gioacchino D., Bianconi A., Wu Z. Y. and
Saini N. L., EPL 90 57001 (2010).\\

\bibitem{BondinoPRL} Bondino F., Magnano E., Malvestuto M., Parmigiani
F., McGuire M. A., Sefat A. S., Sales B. C., Jin R., Mandrus D.,
Plummer E. W., Singh D. J. and Mannella N., Phys.  Rev.  Lett.  101
267001 (2008); \\
Bondino F., Magnano E., Booth C.H., Offi F., Panaccione
G., Malvestuto M., Paolicelli G., Simonelli L., Parmigiani F., McGuire
M. A., Sefat A. S., Sales B. C., Jin R., Vilmercati P., Mandrus D.,
Singh D.J. and Mannella N., Phys.  Rev.  B 82 014529 (2010).\\

\bibitem{KrollPRB} Kroll T., Bonhommeau S.,Kachel T., Duerr H. A.,
Werner J., Behr G.,Koitzsch A., Huebel R.,Leger S., Schoenfelder R.,
Ariffin A. K., Manzke R., de Groot F. M. F., Fink 4, Eschrig
H.,Buechner B., and M. Knupfer, Phys.  Rev.  B 78, 220502
(2008).\\

\bibitem{Ignatov08} Ignatov A., Zhang C. L., Vannucci M., Croft M.,
Tyson T. A., Kwok D., Qin Z., Cheong S.-W., arXiv:0808.2134v2.\\


\bibitem{MustreJS09} Mustre de Leon J., Lezama-Pacheco J., Bianconi A. and
Saini N.L., J. Supercond.  Nov.  Mag.  22, 579-583 (2009).\\

\bibitem{ChangPRB09} Chang B. C., You Y. B., Shiu T. J., Tai M. F., 
Ku H. C., Hsu Y. Y., Jang L. Y., Lee J. F., Wei Z., Ruan K. Q., and 
Li X. G., Phys. Rev. B 80, 165108 (2009).\\

\bibitem{ShenPRB} Yang W. L., Sorini A. P., Chen C-C., Moritz B., Lee
W.-S., Vernay F., Olalde-Velasco P., Denlinger J. D., Delley B., Chu
J.-H., Analytis J. G., Fisher I. R., Ren Z. A., Yang J., Lu W., Zhao
Z. X., van den Brink J., Hussain Z., Shen Z.-X., and Devereaux T. P.,
Phys.  Rev.  B 80, 014508 (2009).\\

\bibitem{BondinoPRB}Parks Cheney C., Bondino F., Callcott T. A.,
Vilmercati P., Ederer D., Magnano E., Malvestuto M., Parmigiani F.,
Sefat A. S., McGuire M. A., Jin R., Sales B. C., Mandrus D., Singh D.
J., Freeland J. W., Mannella N., Phys.  Rev.  B 104518, 81 (2010).\\


\bibitem{MizJPSJ09} Mizuguchi Y., Tomioka F., Tsuda S., Yamaguchi T.,
and Takano Y., J. Phys.  Soc.  Jpn.  78, 074712 (2009).\\

\bibitem{KidaJSPJ09}
Kida, T., Matsunaga, T., Hagiwara, M., Mizuguchi, Y., 
Takano, Y., Kindo, K., 2009. J. Phys. Soc. Jpn. 78, 113701  (2009).\\

\bibitem{BianPRB85} Bianconi A., Fritsch E., Calas G., and Petiau J.,
Phys. Rev.  B 32, 4292 (1985).\\

\bibitem{SeK1} Kvashnina K. O., Butorin S. M., Cui D., Vegelius J.,
Puranen A., Gens R. and Glatzel P., J. of Phys: Conf.  Ser.  190,
012191 (2009).\\

\bibitem{SeK2} Chen C. L., Rao S. M., Dong C. L., Chen J. L., Huang T. W,
Mok B. H., Ling M. C., Wang W. C., Chang C. L., Chan T. S., Lee J.
F., Guo J.-H. and Wu M. K., arXiv:1005.0664v2.

\end{thebibliography}
\end{document}